# Fast Data

Moving beyond from Big Data's map-reduce

Adam Lev-Libfeld, Alexander Margolin

Tamar Labs

Tel-Aviv, Israel

adam@tamarlabs.com

*Abstract*—**Big Data may not be the solution many are looking for. The latest rise of Big Data methods and systems is partly due to the new abilities these techniques provide, partly to the simplicity of the software design and partly because the buzzword itself has value to investors and clients. That said, popularity is not a measure for suitability and the Big Data approach might not be the best solution, or even an applicable one, to many common problems. Namely, time dependent problems whose solution may be bound or cached in any manner can benefit greatly from moving to partly stateless, flow oriented functions and data models. This paper presents such a model to substitute the traditional map-shuffle-reduce models.**

I. INTRODUCTION

The map-reduce model is the de facto standard for processing large amounts of data (mostly known as Big Data). This technique is based on the idea that for every given problem our dataset holds some form of the solution, which can be found by permutation of the data items, and fusion of that permutated data in a way that eliminates all unnecessary information in the data set [1][3]. This set of assumptions is generic by design and was created as such to service the needs of emerging data warehouse concept [2][3]. As often happens once a method gains popularity, there will also be some who will use it without consideration of appropriateness, applicability or even usefulness, same is the case for map-reduce, where companies use the Big Data buzzword to instill trust in clients or investors, and even scare off the competition.

As industry acknowledges these changes, players are moving towards more tightly constrained tools to mitigate the effects of using map-reduce in an inappropriate manner [4], these are usually surface solutions to the underlying problematic assumptions on the nature of the problems they are trying to solve, and the data they hold. The solution proposed in this article is part of a new approach to looking at problems. This approach, although already in use by many (mostly under the names "Stream Processing"[6][7]) has yet to find a leading paradigm and is currently a collection of tools and techniques [4][5][6] more than an architectural model to be used to the benefit of the solution. This article presents one such paradigm.

II. ASSUMPTIONS AND REALITIES

The map-reduce paradigm is based on several assumptions which, in order to satisfy, software engineers go to great lengths in changing their system. These changes might hinder the performance and maintainability of the system as a whole, in a manner that will overcome any advantages the system may have gained from using map-reduce.

In order to deal with these inaccurate assumptions, one must first identify them, three of these assumptions are:

A. *Completeness of Data*

It is assumed that the answer is, or can be derived from the data we have at any given moment.

B. *Independence of Data Set Calculations*

It is assumed that any action performed on the data set is independent from any other in any way, and can be done concurrently with no side effects.

C. *Relevancy Distinguishability*

It is assumed that we can and will distinguish which data is relevant and which is irrelevant to the calculation at any given moment.

Once we call attention to these assumptions, it is easy to see why they were so appealing (and confusing). In the real world we usually face something very similar, but inherently different:

D. *Contextual Completeness*

The data we have is complete only in the context in which it arrived. As our context is time bound, and time is constantly progressing, the information may never be complete and we may never have all the information needed to form a perfect[1]

---

[1] Such answer may be more than we want. One of the ways we overcome this obstacle is by better defining the error margins allowed for an answer to be considered "correct".

answer, or even a correct one.

### E. Partially Dependent Calculation

There exists a *series* of calculations, each takes as input the output of its' predecessor, for which data (input) needed is less dependent in every step. i.e. the further you go down the chain of calculations, the more independent and less state aware each calculation is, as state has been partially resolved by previous calculations.

### F. Emergent Relevancy

In essence the complement of context completeness, it is not only uncertain that we will have complete information to form an answer, but it is also possible that we would not be able to distinguish between several options which would lead to a correct answer and which will not. Having that said, upon trying different answers we can deduce which answer is best, in a manner that will eventually produce the best possible answer (that is - the least irrelevant).

## III. HANDLING UNCERTINTY

In order to deal with these complex realities one might, as many do, erect mechanisms to "clean" the uncertainties out of the system. These mechanisms tend to have significant cost, in either computation time system complexity or both. We suggest a formalized approach that, by taking context and uncertainty into account enables the creation of cleaner (and simpler) system design, which is easier to maintain, extend and scale.

The new approach consists of three main elements: Filter, Splitter and Dehydrator.

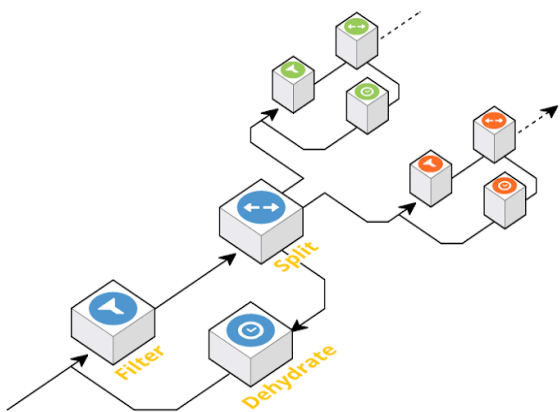

These three elements work in unison to perform the same objective of the map-reduce, filters acting as reducers and splitters as mapping/shuffling operations. The objective of the dehydrator element is to create a controlled feedback loop; effectively prolonging past context so greater meaning could be derived from latest information.

## IV. FILTER

The filter acts as a *reducing* measure to the incoming flow of data, enabling following elements to be more specific as they only have to process data which is already assumed to be (at least somewhat) relevant.

**Stateless Filters**

These can be applied to a single input, or to a whole data array in parallel. This filter's only input is the data element itself, and it has no regard to the results of neighboring filters.

**Aggregation (top X) Filters**

The stateful[2] version of the stateless filter, the input for this filter is both a data element and some aggregation of the previous elements, this aggregation could vary from a simple internal count to a full ordered shared cache. Using several tiers of these filters can minimize shared state; each tier aggregates internally and passes a "soft" decision to the next. This soft output may also vary to consist anything from a numerical result with a notation of margin of error to a set of several elements (top X elements), only one of which is an optimal answer to the next stage of computation.

## V. SPLIT

A Splitter will be used to *map* data elements to their appropriate next stage(s) through the system. This channels some of the data once contained within some form of state of a system to the topology, as the receiving stage can be sure that certain conditions apply by it being invoked.

**Stateless Splitters**

As in the case of the stateless filter, this splitter applies its logic only to the data it takes as input. Based on the results of the logic, the splitter routes the input (or some derivative of it created by the logic) to the next element.

**Aggregation (memory) Splitters**

This splitter has an internal or shared state[2], which is effected by previous inputs or results, and the splitter uses to route results of its logic to the following elements of the system.

---

[2] Exactly how "much" stateful this element will be can be controlled by moving shared state locally, and duplicating the data elements on the "edges" of the shared data-structure, this will lower co-dependency of the different elements in your system and enable higher scalability (in the cost of space over time, of course).

## VI. Dehydrate

Dehydrators are simplistic time machines. They transport data elements that arrived prematurely in terms of their context right to the future where they might be needed, without loading the system while waiting. This concept is achieved by attaching a time-indexed data store to a clock, storing elements as they arrive to the dehydrator and re-introducing them as inputs to the system once a predetermined time period has passed. In order to determine that time period one must take into account three issues:
1. At what point we consider a data element too old and retire it?
2. At what resolution do we want elements to be "retried" as input and how this resolution shift with element age?
3. Are there any exceptions and special circumstances required by the system logic that may change 1, 2 or both for a certain element?

Once these three factors are determined, constructing a scheduling process is as simple as adding a new split process which attaches each data element with it's appropriate dehydration time, and sends it to the data store.

## VII. Use Case : Real Time Geo-Matching

We shall now bring as example the case of a system built for matching "Question" elements tagged with a geospatial position (latitude, longitude and radius) and "Candidates" who are contently report on location. Both new questions and location notification from users are constantly flowing into the system. Once a candidate intersects with a question, the question had to, under some logic, be sent to that user, or if there is more than one user, to the most suitable one. A question could not be sent twice to the same candidate. If there were no candidates within the question active area, but some were on (or near) its edge the system may request a location update from these candidates. These requests are limited in number and frequency.

The original system had all the characteristics we mentioned before, as the client in this specific instance was a start-up, focus was set on speedy proof of concept (POC) and short time to market. The engineer in charge decided on using a centralized DB for the POC and this went on into the final system.

The system suffered of inability to scale and severely reduced performance even in relatively low concurrent user count.

Originally, the system design revolved around two parts:
1. A monolithic database in which all relevant data is stored (see assumption C), and it is assumed that all data needed for a certain response, if it ever arrived, will be available within it (see assumption A)
2. A set of workers managed by a job queue, each serves requests independently as they come in to form a response based on the described DB and send it back to the relevant user (see assumption B).

The inputs to the system, as mentioned above were either Question or Candidate elements; both are arriving constantly and are of varying relevancy, e.g. questions that no longer need an answer, users that are not near any question or have since moved away, and even candidates who, once matched, will not provide an answer (see assumption F).

After analyzing the available system, its problems and challenges, the suggested system structure was as such:

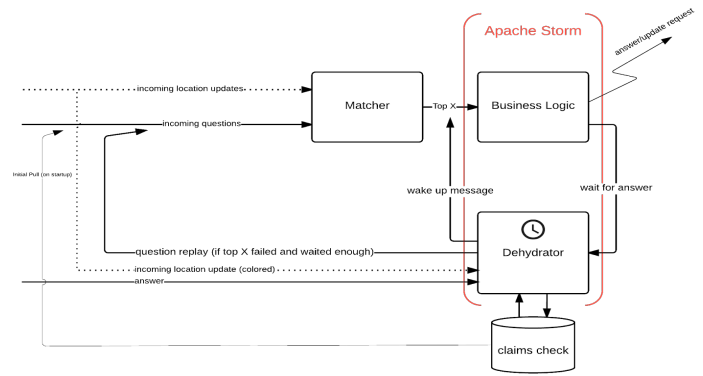

As can be seen in the diagram, this is a fairly straightforward single tier filter-split-dehydrate setup, with both the filter and the splitter in their aggregator variants.

The Matcher element is a cached filter - caching items using an in-memory RTree and produces upon incoming request the corresponding intersection of possible responses, intersecting an incoming question with known user locations and vice versa.

The Business logic is a memory splitter - mostly independent but shares some statistics with fellow splitters, deciding when to send user a question, a request to update location, or to just forward a question event to the dehydrator to be retried on a later date (or even retired altogether).

Holding all that shared data in an in memory database (Redis [10]) is a crucial part of the system implementation. It allowed for the different filter and splitter instances to start independently and only access the state it needed (see assumption E) in minimal latency.

In Addition, the Dehydrator holds questions as the system waits to a user response, and replays/retires each question, based on business logic. This enables for a question's context to be prolonged (see assumption D) with minimal load overhead on the system.

Upon deployment of this solution, effective performance rose by over 9 orders of magnitude with the system at its final setup supporting of millions of requests per second on a single server.

## VIII. Conclusion

The filter-split-dehydrate model, with its stateless, aggregatory, single-tiered and multi-tiered variants is a versatile model that enables a clean software design reminiscent of map-reduce, and adapted to a partial data reality, in which data elements flow into the system in an stochastic fashion. The likes of this model is already being used in the industry to various ends and was proven to, once used correctly to have positive results on performance and scalability.

## Acknowledgments

We would like to thank Vioozer Israel [9] for allowing us to use their system diagram for the use case and data in the course of this work.